\documentstyle[12pt]{article}
\begin{document}
\begin{center}
{\Large \bf Description of the weak interactions within
the framework of electrodynamics}
\bigskip

{\large D.L.~Khokhlov}
\smallskip

{\it Sumy State University, R.-Korsakov St. 2\\
Sumy 244007 Ukraine\\
e-mail: others@monolog.sumy.ua}
\end{center}

\begin{abstract}
The weak interactions are described within
the framework of electrodynamics. The massive vector field
is considered as photon with the effective mass defined in the
second order of the perturbation theory.
Muon and tau-lepton are considered to be
composite particles having the structure
electron-positron-electron.
The mass of muon is defined by the section
of two-photon annihilation.
The mass of tau-lepton is defined by the section
of three-photon annihilation.
\end{abstract}

\section{Introduction}

As known~\cite{B},
the weak interactions can be described by the Lagrangian
\begin{equation}
L\rightarrow -\frac{e^2}{m^2}J_{\mu}(x)J_{\nu}(x)
\rightarrow -G_{F}J_{\mu}(x)J_{\nu}(x).
\label{eq:lwi}
\end{equation}
From this the weak interactions can be considered within the
framework of the theory with the massive vector field
where the mass of the vector field is defined by the
Fermi constant
\begin{equation}
m\sim\left(\frac{e^2}{G_{F}}\right)^{1/2}.
\label{eq:mvf}
\end{equation}
Since the theory with the massive vector fields is
nonrenormalizable, this must follow from some
renormalizable theory.

In the $SU(2)\times U(1)$ electroweak theory~\cite{B},
original bosonic and fermionic fields are massless.
These acquire masses due to the Higgs mechanism.
The scalar field $\varphi$ is introduced, with its potential
is given by
\begin{equation}
V=-\lambda^2(|\varphi|^2-\eta^2)^2.
\label{eq:sf}
\end{equation}
Interaction of the vector fields with the scalar field gives
the masses to the vector fields~$Z$,~$W$
\begin{equation}
m_{Z,W}\sim e\eta.
\label{eq:mvf1}
\end{equation}
Also the massive scalar Higgs field $H$ arises, with its mass
is given by
\begin{equation}
m_{H}=2\lambda\eta.
\label{eq:mhf}
\end{equation}
Masses of the fermions, leptons and quarks, arise due to the
Yukawa interaction
\begin{equation}
f(\bar\psi_{L}\psi_{R}\phi+\bar\psi_{R}\psi_{L}\bar\phi)
\label{eq:yui}
\end{equation}
where the couplings $f$ are different for each fermion.

In the $SU(2)\times U(1)$ theory,
the value of $\eta$ is defined by the Fermi constant
\begin{equation}
\eta\sim\left(G_{F}\right)^{-1/2}.
\label{eq:eta}
\end{equation}
The value of $\lambda$ is not defined.
Yukawa couplings are defined by the masses of the fermions
\begin{equation}
f\sim m_{fer}.
\label{eq:yuc}
\end{equation}
Thus the $SU(2)\times U(1)$ theory do not define the mass of the
Higgs field and do not explain the hierarchy of the leptons and
quarks masses.

\section{Massive vector field}

In the electrodynamics,
in the second order of the perturbation theory,
the transition of photon into itself via the birth and
annihilation of the electron-positron pair occurs~\cite{L}.
At the energy of photon equal to the mass of electron
when the formation
of the real electron-positron pair is possible,
the matrix element of transition
is given by
\begin{equation}
M\sim\frac{\alpha^2}{m_{e}}.
\label{eq:mso1}
\end{equation}
This can be interpreted as that photon acquires an effective mass
\begin{equation}
m\sim\alpha^{-2}m_{e}.
\label{eq:mi}
\end{equation}
In this way we can consider interactions with the massive vector
field where the mass is given by eq. (\ref{eq:mi})
\begin{equation}
L\rightarrow -e^2\frac{\alpha^4}{m_{e}^2}J_{\mu}(x)J_{\nu}(x).
\label{eq:lso}
\end{equation}

Let us assume that the weak interactions can be described
within the framework of electrodynamics.
$Z$-,~$W$-bosons can be considered as photons with effective masses
defined in the second order of the perturbation theory.
Then the weak interactions can be described by the Lagrangian
(\ref{eq:lso}). The theory with the Lagrangian (\ref{eq:lso})
is renormalizable, since this emerges
within the framework of electrodynamics.

Consider the process of scattering of photon by photon proceeding
in the second order of the perturbation theory
\begin{equation}
\gamma\gamma\rightarrow e^{-}e^{+}e^{-}e^{+}\rightarrow\gamma\gamma.
\label{eq:spp}
\end{equation}
Determine the effective mass of photon from the section of the
reaction (\ref{eq:spp}).
At $\hbar\omega<<m_{e}c^2$,
the section of scattering for non-polarized photons
grows as~\cite{L}
\begin{equation}
\sigma=0.031\alpha^2r_{e}^2
\left({\hbar\omega}\over{m_{e}c^2}\right)^6.\label{eq:j1}
\end{equation}
At $\hbar\omega>>m_{e}c^2$,
the section of scattering for non-polarized photons
diminishes as~\cite{L}
\begin{equation}
\sigma=4.7\alpha^4\left(\frac{c}{\omega}\right)^2.
\label{eq:j2}
\end{equation}
In the limiting case $\hbar\omega={m_{e}c^2}$ when the formation
of the real electron-positron pair is possible,
the section of scattering is
\begin{equation}
\sigma=0.031\alpha^2r_{e}^2.
\label{eq:ssc}
\end{equation}
This gives the radius of interaction
\begin{equation}
r=\left({0.031\over\pi}\right)^{1/2}\alpha r_{e}
\label{eq:rsc}
\end{equation}
and the effective mass of photon
\begin{equation}
m=\left(\frac{\pi}{0.031}\right)^{1/2}\frac{\hbar}{\alpha r_{e}c}=
\left(\frac{\pi}{0.031}\right)^{1/2}\frac{m_{e}}{\alpha^2}.
\label{eq:msc}
\end{equation}
According to eqs. (\ref{eq:rsc}), (\ref{eq:msc}),
the radius of interaction is $r\sim 2\cdot 10^{-16} {\rm \ cm}$,
and the effective mass of photon is
$m=96.5\cdot 10^2 {\rm \ GeV}$.
This value is close to the mass of Z-boson equal
to $m_Z=93\cdot 10^2 {\rm \ GeV}$~\cite{R}.

\section{Mass hierarchy between lepton generations}

Let us consider the problem of mass hierarchy between lepton
generations within the framework of electrodynamics.
The mass of electron is a parameter defined from the
experimental data. The theory have to define the masses
of muon and tau-lepton.

Let us assume that muon and tau-lepton are composite particles.
Namely muon and tau-lepton have the following structure
\begin{equation}
\mu^-\equiv e^-e^+e^-\label{eq:a}
\end{equation}
\begin{equation}
\tau^-\equiv e^-e^+e^-.\label{eq:b}
\end{equation}
Muon arises due to the reaction
\begin{equation}
e^- +2\gamma\rightarrow e^-e^+e^-,\label{eq:c}
\end{equation}
and tau-lepton arises due to the reaction
\begin{equation}
e^- +3\gamma\rightarrow e^-e^+e^-.\label{eq:d}
\end{equation}

The mass of muon is defined by the energy required to bear
electron-positron pair in the process of two-photon annihilation.
Since two-photon annihilation is characterized by the classical
radius of electron $r_e$, the mass of muon is of order
\begin{equation}
m_{\mu}\sim{\hbar\over{cr_e}}.\label{eq:g}
\end{equation}
According to eq.~(\ref{eq:g}), the mass of muon is
$m_{\mu}\sim 70\ {\rm MeV}$ that agrees with the experimental
value $m_{\mu}=106\ {\rm MeV}$~\cite{R}.

The mass of tau-lepton is defined by the energy required to bear
electron-positron pair in the process of three-photon annihilation.
Let us determine the radius which characterizes three-photon
annihilation.
The section of two-photon annihilation in the non-relativistic
limit is given by~\cite{L}
\begin{equation}
\bar\sigma_{2\gamma}=\pi\left(e^2\over{m_{e}c^2}\right)^2{c\over v}.
\label{eq:h}
\end{equation}
The section of three-photon annihilation in the non-relativistic
limit is given by~\cite{L}
\begin{equation}
\bar\sigma_{3\gamma}={4(\pi^2-9)\over 3}\alpha
\left(e^2\over{m_{e}c^2}\right)^2{c\over v}.\label{eq:j}
\end{equation}
From this the radius which characterizes three-photon
annihilation is
\begin{equation}
r_{3\gamma}=r_e\left({4(\pi^2-9)\over{3\pi}}\alpha\right)^{1/2}.
\label{eq:k}
\end{equation}
Then the mass of tau-lepton is of order
\begin{equation}
m_{\tau}\sim{\hbar\over{cr_{3\gamma}}}.\label{eq:k1}
\end{equation}
According to eq.~(\ref{eq:k1}), the mass of tau-lepton is
$m_{\tau}\sim 2200\ {\rm MeV}$ that agrees with the experimental
value $m_{\tau}=1784\ {\rm MeV}$~\cite{R}.

In order to describe the decays of muon
and tau-lepton
\begin{equation}
\mu^-\rightarrow e^- +\bar\nu_{e}\nu_{\mu}\label{eq:s1}
\end{equation}
\begin{equation}
\tau^-\rightarrow e^- +\bar\nu_{e}\nu_{\tau}\label{eq:s2}
\end{equation}
within the framework of electrodynamics
let us assume that the following reactions occur
\begin{equation}
\gamma\rightarrow\bar\nu_{e}\bar\nu_{e}
\label{eq:gnu}
\end{equation}
\begin{equation}
\bar\nu_{e}+\gamma\rightarrow\nu_{\mu}
\label{eq:x}
\end{equation}
\begin{equation}
\bar\nu_{e}+2\gamma\rightarrow\nu_{\tau}.
\label{eq:y}
\end{equation}
Then combining eqs.~(\ref{eq:a}), (\ref{eq:c}),
(\ref{eq:gnu}), (\ref{eq:x})
we obtain the reaction
for the decay of muon
\begin{equation}
\mu^-\equiv e^-e^+e^-\rightarrow e^- +2\gamma
\rightarrow e^- +\gamma\bar\nu_{e}\bar\nu_{e}
\rightarrow e^- +\bar\nu_{e}\nu_{\mu}\label{eq:dm}.
\end{equation}
Combining eqs.~(\ref{eq:b}), (\ref{eq:d}),
(\ref{eq:gnu}), (\ref{eq:y})
we obtain the reaction
for the decay of tau-lepton
\begin{equation}
\tau^-\equiv e^-e^+e^-\rightarrow e^- +3\gamma
\rightarrow e^- +2\gamma\bar\nu_{e}\bar\nu_{e}
\rightarrow e^- +\bar\nu_{e}\nu_{\tau}\label{eq:dt}.
\end{equation}

\end{document}